\documentclass[twocolumn,aps,superscriptaddress,10pt,prapplied,showkeys]{revtex4-2}

\UseRawInputEncoding

\usepackage[english]{babel}

\usepackage{graphicx}
\usepackage{bm}
\usepackage[dvipsnames]{xcolor}
\usepackage[mathlines]{lineno}
\usepackage{braket}
\usepackage{tabularx}
\usepackage{amsmath}
\usepackage{mathtools}
\usepackage[noend]{algpseudocode}
\usepackage{algorithm}
\usepackage{ulem}

\usepackage{hyperref}

\begin{document}

\title{Broadband Fourier transform spectroscopy of quantum emitters photoluminescence \\ with sub-nanosecond temporal resolution}

\author{Issam Belgacem}
\affiliation{ 
Institute of Photonics and Quantum Sciences, SUPA, Heriot-Watt University,
Edinburgh EH14 4AS, UK
}

\author{Pasquale Cilibrizzi}
\affiliation{ 
Institute of Photonics and Quantum Sciences, SUPA, Heriot-Watt University,
Edinburgh EH14 4AS, UK
}

\author{Muhammad Junaid Arshad}
\affiliation{ 
Institute of Photonics and Quantum Sciences, SUPA, Heriot-Watt University,
Edinburgh EH14 4AS, UK
}

\author{Daniel White}
\affiliation{ 
Institute of Photonics and Quantum Sciences, SUPA, Heriot-Watt University,
Edinburgh EH14 4AS, UK
}

\author{Malte Kroj}
\affiliation{ 
Institute of Photonics and Quantum Sciences, SUPA, Heriot-Watt University,
Edinburgh EH14 4AS, UK
}

\author{Christiaan Bekker}
\affiliation{ 
Institute of Photonics and Quantum Sciences, SUPA, Heriot-Watt University,
Edinburgh EH14 4AS, UK
}

\author{Margherita Mazzera}
\affiliation{ 
Institute of Photonics and Quantum Sciences, SUPA, Heriot-Watt University,
Edinburgh EH14 4AS, UK
}

\author{Brian Gerardot}
\affiliation{ 
Institute of Photonics and Quantum Sciences, SUPA, Heriot-Watt University,
Edinburgh EH14 4AS, UK
}

\author{Angelo C. Frangeskou}
\affiliation{ 
Department of Physics, University of Warwick, Coventry CV4 7AL, United Kingdom
}

\author{Gavin W. Morley}
\affiliation{ 
Department of Physics, University of Warwick, Coventry CV4 7AL, United Kingdom
}

\author{Nguyen Tien Son}
\affiliation{Department of Physics, Chemistry and Biology, Link\"oping University, SE-581 83 Link\"oping, Sweden}

\author{Jawad Ul-Hassan}
\affiliation{Department of Physics, Chemistry and Biology, Link\"oping University, SE-581 83 Link\"oping, Sweden}

\author{Takeshi Ohshima}
\affiliation{Quantum Materials and Applications Research Center (QUARC), National Institutes for Quantum Science and Technology (QST), 1233Watanuki, Takasaki, Gunma 370-1292, Japan}
\affiliation {Department of Materials Science, Tohoku University, Aoba, Sendai, Miyagi 980-8579, Japan}

\author{ Hiroshi Abe}
\affiliation{Quantum Materials and Applications Research Center (QUARC), National Institutes for Quantum Science and Technology (QST), 1233Watanuki, Takasaki, Gunma 370-1292, Japan}

\author{Lorenzo Vinco}
\affiliation{ 
Department of Physics, Politecnico di Milano, P.zza L. da Vinci 32, 20133 Milano, Italy
}

\author{Dario Polli}
\affiliation{ 
Department of Physics, Politecnico di Milano, P.zza L. da Vinci 32, 20133 Milano, Italy
}
\affiliation{CNR-Institute for Photonics and Nanotechnologies (CNR-IFN), Milan 20133, Italy}

\author{Giulio Cerullo}
\affiliation{ 
Department of Physics, Politecnico di Milano, P.zza L. da Vinci 32, 20133 Milano, Italy
}
\affiliation{CNR-Institute for Photonics and Nanotechnologies (CNR-IFN), Milan 20133, Italy}

\author{Cristian Bonato}
\email {c.bonato@hw.ac.uk}
\affiliation{ 
Institute of Photonics and Quantum Sciences, SUPA, Heriot-Watt University,
Edinburgh EH14 4AS, UK
}

\begin{abstract}
The spectral characterization of quantum emitter luminescence over broad wavelength ranges and fast timescales is important for applications ranging from biophysics to quantum technologies.
Here we present the application of time-domain Fourier transform spectroscopy, based on a compact and stable birefringent interferometer coupled to low-dark-count superconducting single-photon detectors, to the study of quantum emitters. We experimentally demonstrate that the system enables spectroscopy of quantum emitters over a broad wavelength interval from the near-infrared to the telecom range, where grating-based spectrometers coupled to InGaAs cameras are typically noisy and inefficient. We further show that the high temporal resolution of single-photon detectors, which can be on the order of tens of picoseconds, enables the monitoring of spin-dependent spectral changes on sub-nanosecond timescales.

\end{abstract}

\maketitle

\section {Introduction}

Photo-luminescence (PL) spectroscopy of optical emitters (single or small ensembles) is a task of primary importance in different fields. 
Single-molecule detection and spectroscopy in chemistry and biology, either label-free or through the use of fluorescent markers, can reach a sufficient spatial resolution to non-destructively monitor biochemical processes at the subcellular level \cite{koveal_functional_2024, maiti_structural_2023, wang_fluorescent_2023,cubeddu_time-resolved_2002, gobets_time-resolved_2001}. Applications range from fundamental studies in life sciences to biomedical diagnostics. In quantum technology, PL spectroscopy of single quantum emitters, such as single atoms, molecules, impurities/defects, or nanostructures in solid-state materials, is essential to characterize their basic electronic level structure and excited-state dynamics and to implement qubits for quantum communication and sensing. 

In many of these applications, PL spectroscopy is crucial to determine the interaction between optical transitions and the spin structure of the ground and excited states, to create a spin-photon interface. Recent experiments have proven the basic principles of quantum networks that exploit spin-photon interfaces associated with quantum dots in III-V semiconductors \cite{gao_observation_2012, javadi_spinphoton_2018, gangloff_quantum_2019}, point defects in several materials such as diamond \cite{hermans_qubit_2022,parker_diamond_2024, knaut_entanglement_2024, stolk_metropolitan-scale_2024}, silicon carbide \cite{koehl_room_2011, christle_isolated_2015, christle_isolated_2017, nagy_high-fidelity_2019, cilibrizzi_ultra-narrow_2023} and silicon \cite{redjem_single_2020, durand_broad_2021, inc_distributed_2024}, rare-earth ions in solids \cite{wu_near-infrared_2023, chen_parallel_2020, gritsch_optical_2024}, or molecular spin qubits \cite{bayliss_optically_2020, mena_room-temperature_2024}. In particular, emission in the telecom range, e.g. O-band  (1260-1360nm) and C-band (1530-1565nm), is privileged for applications requiring long-distance sharing of photonic quantum states \cite{awschalom_quantum_2018, ecker_quantum_2024}. Electronic spins associated with single-point defects, in particular the nitrogen-vacancy (NV) centers in diamond, are also widely utilized as a localized quantum sensor to probe magnetic and electric fields, as well as temperature at the nanoscale \cite{aslam_quantum_2023, neuling_prospects_2023, budakian_roadmap_2024, rovny_new_2024}. Although the operating wavelength is typically less important in these applications, there are cases where specific wavelengths might be preferable to avoid interference with the properties of the sample under investigation \cite{chen_noninvasive_2024}.  

Spectroscopic characterization of quantum emitters is typically performed in the frequency domain, using a dispersive optical element, such as a diffraction grating, coupled to a multichannel detector, such as a CCD camera or a detector array. Alternatively, it can be performed in the time domain using Fourier-transform (FT) spectroscopy. FT spectroscopy utilizes an interferometer to create two replicas of an optical waveform and records their interference as a function of the replicas delay, the so-called interferogram, whose FT as a function of delay provides the spectrum \cite{jacquinot_new_1960, bates_fourier_1978, vij_handbook_2006}. In the realm of quantum emitters, FT PL spectroscopy has been used to measure the linewidth and decoherence properties of single quantum dots \cite{kammerer_interferometric_2002, zwiller_single-photon_2004, kuroda_single-photon_2007, adachi_decoherence_2007, holmes_spectral_2015}, carbon nanotubes \cite{korlacki_optical_2007} and defects in diamond \cite{marshall_coherence_2011}. 

Here, we demonstrate broadband FT spectroscopy of single quantum emitters, using a compact and stable common-path birefringent interferometer coupled to superconducting nanowire single-photon detectors (SNSPDs). We provide a quantitative numerical and experimental demonstration of the advantages of this approach through spectroscopic measurements of two quantum emitters, —the divacancy in SiC and the NV center in diamond,— emitting in the near-infrared and telecom wavelength ranges, where high-performance silicon-based detectors cannot operate. We further show that, given the high temporal resolution of single-photon detectors, our approach can be used to detect spectral changes with sub-nanosecond temporal resolution. To demonstrate this, we perform spin-selective time-resolved measurements of NV centers, showing how this approach enables the monitoring of spin-resolved dynamics, which is essential for the implementation of spin-photon interfaces.

\section {Simulations}
\label{sec:simulations}

In this first section of the work we quantitatively compare, through numerical simulations, the performance of frequency-domain grating-based spectroscopy with time-domain FT spectroscopy for the case of a single quantum emitter operating in the wavelength range of InGaAs cameras.
A grating spectrometer consists of a dispersive element, such as a linear diffraction grating, which separates the wavelengths into different wave-vectors, which are then projected onto a detector array, such as a CCD camera. The performance of a camera for spectroscopy applications is determined by several parameters. The \textit{quantum efficiency} $\eta (\lambda)$ represents the fraction of incident photons converted to photo-electrons (dependent on the photon wavelength $\lambda$). The \textit{gain} $G$ (expressed in photo-electrons/count) describes how many photo-electrons are required to generate one detector count. The \textit{dark noise} $N_d$ (electrons/pixel/second) corresponds to the number of thermally-excited background electrons (i.e. not originating from an incident optical photon). The \textit{readout noise} $R$ describes the noise of the electronic readout chain, expressed as the number of electrons per readout shot. This noise source is only active when the electronic signal for the pixel is measured, and is therefore independent of the integration time. Finally, the \textit{full well capacity} $W$ corresponds to the number of photo-electrons that can be stored for each pixel, before saturating.

Given an input photon spectrum $N_{ph} (\lambda)$ defined as the number of photons per second for each wavelength $\lambda$ and an integration time $\Delta t$, the mean number of photo-electrons $\overline{N}_e (\lambda)$ in a given pixel can be computed as \cite{konnik_high-level_2014}:

\begin{equation}
 \overline{N}_e (\lambda) = \left[ \eta (\lambda) \cdot N_{ph} (\lambda) + N_d \right]  \cdot \Delta t + R
\end{equation}

The number of photo-electrons $N_e (\lambda)$ for each pixel can be described as a Poisson process with mean $ \overline{N}_e (\lambda)$. If $N_e (\lambda) > W$, the well capacity is saturated and $N_e (\lambda)$ can be set to $N_e (\lambda) = W$, posing a limit to the increase in integration time to improve the signal-to-noise ratio (SNR) of the obtained spectrum. The number of photo-electrons $N_e (\lambda)$ is then converted to counts by dividing it by the gain $G$. In our simple model, we neglect nonlinearities of the detection chain and quantization noise.

An FT spectrometer consists of an amplitude division interferometer (e.g. Michelson or Mach-Zehnder), which splits the input optical beam into two replicas and recombines them after a tuneable delay to create an interferogram \cite{padgett_static_1995, thyrhaug_single-molecule_2019}. The interferogram is then Fourier-transformed with respect to the time delay $\tau$  between the replicas to obtain the frequency spectrum of the input optical field, following the Wiener-Khinchin theorem. Here the main parameters determining the performance of spectrum estimation are the detector quantum efficiency $\eta (\lambda)$ and the dark count rate $N_d$. 

The number of counts detected by the FT spectrometer can be written as a Poisson process with mean $N_C^{(F)}$ given by

\begin{equation}
    N_C^{(F)} (\tau) = \int_{-\infty}^{+\infty} e^{i \omega \tau} \eta (\omega) N_{ph} (\omega) d\omega + N_d
\end{equation}
where $\omega = 2\pi c / \lambda$. In our simulation we assume the detector saturates at $W_F = 5 \times 10^6$ counts per second (cps), i.e. if $N_C^{(F)} (\tau) > W_F$ then we set $N_C^{(F)} (\tau) = W_F$.

\begin{figure*}[!htbp]
\centering
\includegraphics[width=1\textwidth]{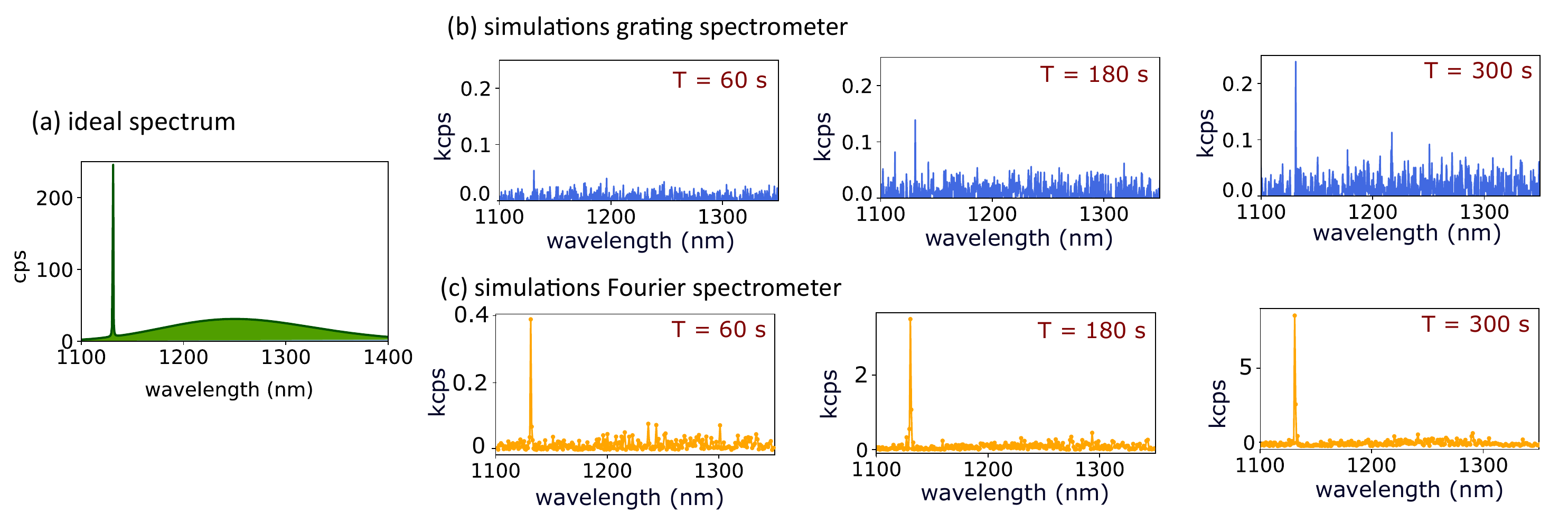}
\caption{\textbf{Numerical simulations of single-emitter PL comparing a grating and FT spectrometer.}  We consider a quantum emitter with Debye-Waller factor 0.04 and brightness $2$kcps, with the spectrum shown in \textbf{(a)} (corresponding to a divacancy in 4H SiC \cite{falk_polytype_2013}). Simulated spectra measured with a grating spectrometer coupled to a InGaAs camera ($G = 75 e^{-}$/count, $N_d = 3.2 \text{k}e^{-}$/count, $R = 400 e^{-}$, $W = 4.5 \text{M}e^{-}$) are shown in blue in \textbf{(b)} (total integration time $T = 60$ s, $T = 180$ s, $T = 300$ s from left to right, respectively). The corresponding FT spectra ($\eta = 0.8$, $N_d = 50$ cps) are shown in orange in  \textbf{(c)}, for the same total integration times as in (b). We take the same spectral resolution for both spectrometers.}
\label{fig:simulations}
\end{figure*}

FT spectroscopy entails several potential advantages. First of all, all wavelengths are multiplexed on the same pixel, resulting in a better SNR due to the higher total intensity, as SNR goes as $1/\sqrt{N}$ for $N$ detected photons (Fellget's advantage \cite{fellgett_ultimate_1949}). Second, FT instruments do not require entrance/exit slits to achieve high spectral resolution, resulting in significantly higher throughput than grating spectrometers, especially for low spatial coherence sources (Jacquinot advantage \cite{jacquinot_luminosity_1954, jacquinot_new_1960}). Third, in the telecom spectral region ($1000-1500$ nm), a crucial range for fiber-based quantum networking applications~\cite{awschalom_quantum_2018,ecker_quantum_2024}, grating spectrometers need to operate with InGaAs cameras which, as outlined above, suffer from high dark noise and consequently lower signal-to-noise ratios~\cite{klem_mesa-isolated_2009, fathipour_advances_2016, zhu_review_2024}. On the other hand, an FT spectrometer can operate with commercially available single-pixel SNSPDs, which feature high detection efficiency (up to almost unity \cite{reddy_superconducting_2020, hu_detecting_2020}) and extremely low dark count rates (potentially down to $10^{-4}$ cps \cite{zhang_ultra-low_2011, shibata_ultimate_2015}). The output of an FT spectrometer can be further split into different spectral components detected with multiple SNSPDs, each optimized for the corresponding spectral range, resulting in broadband spectroscopy. Currently, high-performance SNSPDs are available in a wide wavelength range spanning from the visible to the mid-infrared \cite{lau_superconducting_2023, taylor_low-noise_2023}. Finally, single-photon detectors output short electrical pulses that possess information about photon arrival time. This enables time-resolved spectroscopy on timescales limited only by the detector jitter, which can currently be as low as a few picoseconds \cite{korzh_demonstration_2020, esmaeil_zadeh_efficient_2020}.

The main disadvantage of an FT spectrometer is its single-pixel operation, which results in the need to perform sequential measurements by scanning the delay between the two replicas to extract the light spectrum. This is generally much slower than parallel imaging of different spectrally resolved wave-vectors onto a camera or detector array. Spectral resolution is set by the maximum path delay and can be controlled by the user up to the maximum excursion of the delay line, with a trade-off between spectral resolution and data acquisition time.

To compare the performance of frequency-domain grating-based and time-domain FT spectrometers, we apply numerical simulations based on standard commercial equipment. We consider an emitter with a zero-phonon line (ZPL) at 1131nm (Lorentzian function, 300 MHz linewidth), and with phonon sidebands (PSB) described by a Gaussian function with a linewidth of 20 THz (spectrum in Fig \ref{fig:simulations}(a)). We assume a ratio of 0.04 between the intensity emitted in the ZPL and PSB (Debye-Waller factor), and a total brightness of 2000 cps. These values correspond to typical parameters for a divacancy center in SiC \cite{koehl_room_2011, falk_polytype_2013}. 

For the grating-based spectrometer, we assume that the spectral region 1050-1375 nm is dispersed into 1024 pixels (with a spectral resolution of 0.32 nm), and use values for a commercial InGaAs camera (TeleDyne Princeton Instruments PyLoN-IR:1024, ``high-gain'' settings): gain $G = 75 e^{-}$/count, dark noise $N_d = 5.7 \text{k}e^{-}$/s, readout noise $R = 400 e^{-}$, well-capacity $W = 4.5 \text{M}e^{-}$. We do not include optical losses associated with the entrance slit and the finite grating efficiency. For the FT spectrometer, we consider typical values for the commercial SNSPDs operating in our lab, with $N_d \sim 50$ counts/s. The typical detector response is approximately linear up to a few million counts per second. For both devices, we assume that the detection efficiency is independent of wavelength in the spectral region of interest, with $\eta (\lambda) = \eta = 0.8$.

Fig. \ref{fig:simulations}(b) (blue) and Fig. \ref{fig:simulations}(c) (orange) respectively show simulated spectra for grating-based and FT spectrometers (1, 3, 5 minutes integration time). It clearly demonstrates that the FT spectrometer can identify a single quantum emitter in a few minutes' measurement time, while the grating-based spectrometer has insufficient sensitivity for this purpose over such a timescale. Further simulations are presented in the Supplementary information, comparing the performance of the two techniques for weaker (500 cps, Fig. S3) and brighter (10 kcps, Fig. S2) single quantum emitters. We further explore the impact of the SNSPD dark counts (Fig. S1), showing that FT spectroscopy is very robust, enabling the identification of a single quantum emitter even when the dark count rate is as high as the signal rate.

\begin{figure*}[!htbp]
\centering
\includegraphics[width=1\textwidth]{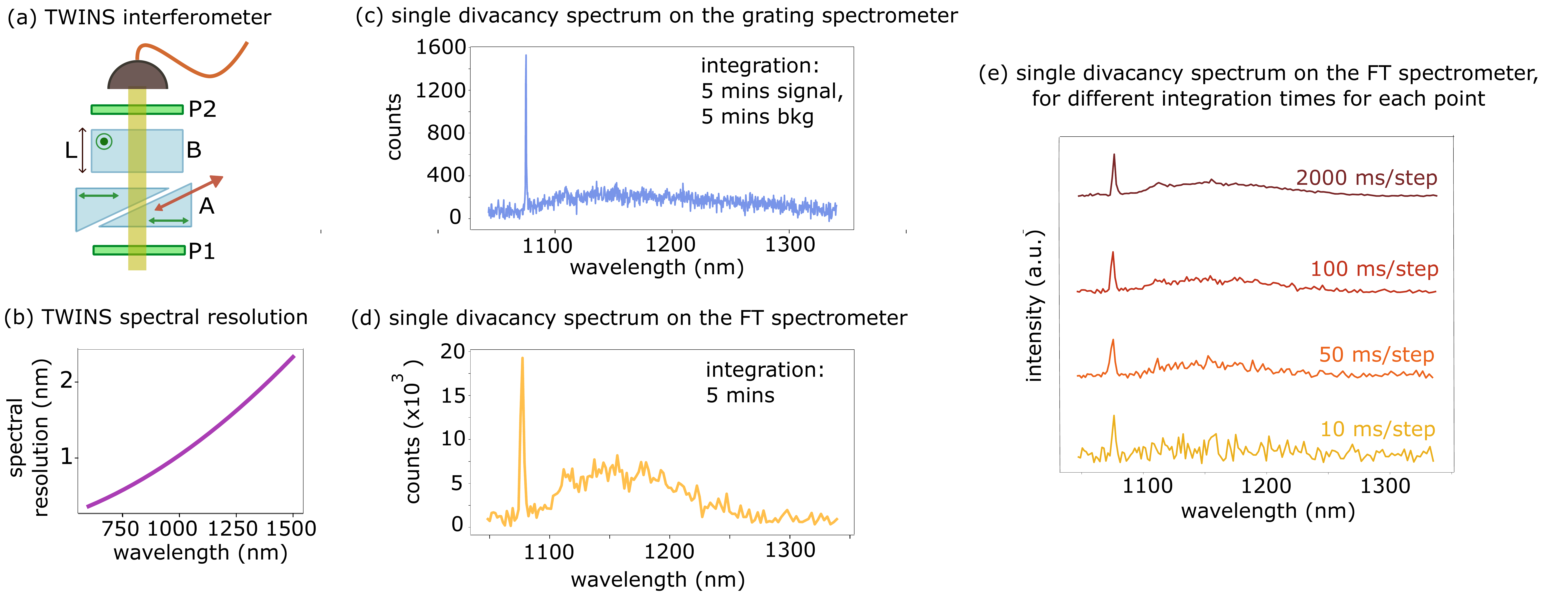}
\caption{\textbf{PL spectroscopy of a single divacancy center in 4H-SiC.} (a) sketch of the TWINS interferometer, consisting of two sliding wedges of a birefringent material (``A'') that create a controllable delay between two replicas of the input optical waveform. Polariser P1 sets the input polarization, while polarizer P2 enables interference of the two orthogonally-polarized replicas. (b) spectral resolution of the TWINS device (GEMINI.HP.ASY.L, NIREOS) used in our experiment. The spectral resolution $\Delta \lambda$ at wavelength $\lambda$ is given by $\Delta \lambda = 0.605 \lambda^2/(\Delta n \cdot x_{max} \cdot \sin\alpha)$ \cite{perri_visible_2021}, where $x_{max}$ is the maximum motor travel range and $\alpha$ is the angle of the wedge. (c) PL spectrum of a single divacancy in 4H-SiC acquired with the grating-based spectrometer coupled to a liquid-nitrogen-cooled InGaAs camera. Observing the spectrum required a background trace to be taken and subtracted from the signal trace (both with 5 minute integration time). The background should at minimum be taken with the excitation laser off, but provides improved SNR when including background fluorescence through collecting emitted light away from the divacancy. (d) Fourier PL spectrum for the same divacancy centre as in (c) acquired with the TWINS interferometer (5 minutes total integration time). No background subtraction is required here. The difference in the shape of the phonon sidebands originates from the difference in spectral resolution (see SI Fig. S4). (e) Fourier PL spectra for the same divacancy center as in (c)/(d), taken at different integration times (1293 time delay steps): 2000 ms/step (total 45 minutes), 100 ms/step (total 153 seconds), 50 ms/step (total 86 seconds) and  10 ms/step (total 37 seconds). The total integration time does not scale linearly with the integration per step, as there is a constant temporal overhead related to motor movement. It is possible to identify the single divacancy, from its ZPL wavelength, even with an integration time on the order of few tens of seconds.}
\label{fig:SiC_Divacancy}
\end{figure*}

\section {Fourier spectroscopy of single near-infrared quantum emitters}
\label {sec:divacancy_experiments}

In this section we experimentally verify the outcomes of the numerical simulations presented in Section \ref{sec:simulations} on a single (hk) divacancy in 4H-SiC. Divacancies in SiC are point defects consisting of neighboring silicon and carbon vacancies (V$_{Si}$V$_{C}$). They exhibit PL in the near-infrared, and a $S=1$ electronic spin, which has been used for experiments on spin-photon interfacing \cite{christle_isolated_2015, christle_isolated_2017, anderson_five-second_2022, bourassa_entanglement_2020} for quantum networking and quantum sensing. In the 4H polytype of SiC, a divacancy can occur in four possible lattice sites, each with specific ZPL wavelengths: 1133 nm for the (hh) divacancy (PL1), 1132nm for the (kk) divacancy (PL2), 1108nm and 1078 nm, respectively, for the two basal divacancies (PL3 and PL4) \cite{falk_polytype_2013}, with additional centers of unknown origins, possibly also associated with divacancies, with ZPLs around 1040nm \cite{he_robust_2024}.

We compare the performances of  a grating spectrometer and a FT spectrometer in detecting the spectrum of a single divacancy in SiC. The grating spectrometer (Acton Research Corporation, SpectraPro-500i) includes a 300 g/mm grating, and is coupled to a liquid nitrogen cooled InGaAs camera. The FT spectrometer consists of a common-path birefringent interferometer \cite{oriana_scanning_2016, preda_linear_2017, perri_excitation-emission_2017, preda_time-domain_2018, perri_hyperspectral_2019, ghosh_broadband_2021}, based on the translating-wedge-based identical pulses encoding system (TWINS) design, sketched in Fig. \ref{fig:SiC_Divacancy} (a). To understand the operation of TWINS, we consider an optical waveform entering a birefringent plate of thickness $L$ with polarization rotated by 45 degrees with respect to the ordinary and extraordinary axes. At the output of the plate, the waveform splits into two replicas with perpendicular polarizations and delay $\tau$ proportional to the plate thickness. The plate (labeled ``A'' in Fig. \ref{fig:SiC_Divacancy}(a)) is cut into a pair of wedges, of which one is transversely translated, to continuously vary $L$ and $\tau$. To access the zero-time delay between the pulses, a birefringent plate (``B'' in Fig. \ref{fig:SiC_Divacancy}(a)), with optical axis rotated by 90 degrees with respect to the wedges, is introduced. Finally, the two delayed replicas are projected to a parallel polarization by a linear polarizer (``P2'' in Fig. \ref{fig:SiC_Divacancy}(a)), allowing their interference on the detector to measure an interferogram. Since the two delayed replicas follow the same beam path, any vibration or environmental fluctuation equally affects them both and is thus canceled, guaranteeing exceptional delay stability and reproducibility.

\begin{figure*}[!hbtp]
\centering
\includegraphics[width=1\textwidth]{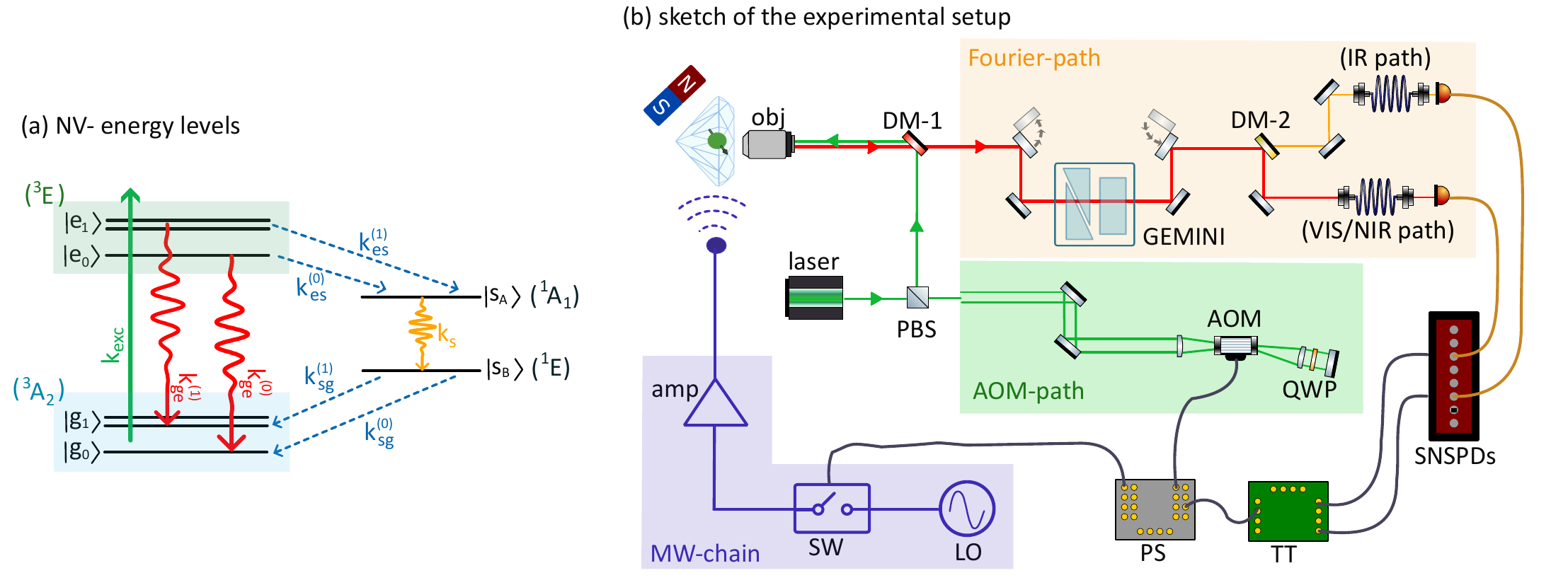}
\caption{\textbf{Broadband time-resolved spectroscopy of NV centers in diamond. }(a) Energy levels for the negatively-charged state of the NV center in diamond.  Radiative (non-radiative) transitions are depicted with solid (dashed) lines and the corresponding rates $\kappa_{lm}^{(j)}$, associated with spin state $\ket{j}$, are described in Section \ref{sec:broadband} of the main text. (b) Sketch of experimental setup for the experiments on NV centers in diamond. The NV centers are addressed with a high-NA objective ('OBJ') in a custom confocal microscope, with the sample mounted on a 3-axes motorized translation stage. Optical excitation is performed by a 532nm laser, either directly or through an acousto-optic modulator (AOM) in double-pass configuration ("AOM-path") with a quarter-wave plate (QWP) in the path to ensure backward propagating light is directed to the sample and not back into the laser by the polarizing beam-splitter (PBS). The PL, separated from the excitation light by a dichroic mirror (DM-1, cut-off wavelength 561 nm) is directed through the GEMINI interferometer (``Fourier-path"), then spectrally split by a second dichroic (DM-2, cut-off wavelength 950 nm), separating the visible and near-infrared components, which are detected by two separate SNSPDs optimized for the respective wavelength ranges. Microwave pulses (``MW-chain'') are created by a switch (SW) acting on a CW tone generated by a microwave oscillator (LO), amplified ('amp') and delivered to the sample by a thin microwave wire. Control pulses for the acousto-optic modulator and the microwave switch are generated by a programmable pulse generator ('PS', Swabian Instruments Pulse Streamer 8/2), and pulses from the SNSPDs are time-tagged by a Swabian Instruments Time Tagger ('TT'). A magnetic field of about 35 gauss is precisely aligned to one of the NV axes by a permanent magnet mounted on a motorized translational stage.  The setup is described in more details in Supplementary Information, section S3.}
\label{fig:NV_levels_setup}
\end{figure*}

We utilize a commercial TWINS interferometer (model GEMINI.HP.ASY.L from NIREOS s.r.l.). The birefringent blocks are made of $\alpha$-barium borate ($\alpha$-BBO), which is transparent from the ultraviolet to 3 $\mu$m wavelength. Block B is a plate with 4.53-mm thickness, while block A is shaped in the form of two wedges with the same apex angle (10$^{\circ}$), one of which is mounted on a motorized translation stage with maximum travel range of 28 mm, which corresponds to a maximum delay between the replicas of 2000 fs at a wavelength of 600nm.  The FT interferometer is combined with an SNSPD (SingleQuantum EOS), with $80\%$ efficiency along one linear polarisation, about $50$ dark counts per second, and about $100$ ps temporal jitter. In the following, we refer to the first method as the `InGaAs Spectrometer' and the second method as the `FT Spectrometer'. 

To compare the two approaches, we investigate an isolated divacancy in 4H-SiC. To create isolated divacancies, the sample was irradiated with 2 MeV electrons at a dose $10^{12}$cm$^{-2}$ and annealed in nitrogen gas flow for 30 minutes at 750 degrees. The sample is cooled to T = 4K in a closed-cycle cryostat and investigated with a home-built confocal microscope, including a high numerical aperture objective (Olympus LCPLN100X-IR, NA = 0.85) held at room temperature inside the cryostat vacuum (see Cilibrizzi et al ~\cite{cilibrizzi_ultra-narrow_2023} for more details about the setup). We excite a single divacancy with a 915nm laser diode, and a 780nm repump laser to stabilize the neutral charge state, detecting approximately 15 kcps on the SNSPD, bypassing the GEMINI interferometer. We then measure its PL spectrum first with a grating-based spectrometer and subsequently with the FT spectrometer. The two spectra are reported in Fig. ~\ref{fig:SiC_Divacancy}.

With the InGaAs spectrometer, we first acquire a background measurement to compensate for the dark noise, approximately 140 counts/(pixel ·s) over the 880 pixels used, with a spectral resolution of 0.3 nm at 1100 nm. A signal measurement is then recorded with an integration time of 5 minutes, which is subtracted from the background trace to yield the optical spectrum, as shown in Fig.\ref{fig:SiC_Divacancy} (c). The visible ZPL at 1078 nm, measured as 1077.96 $\pm$ 0.01 nm from the Lorentzian fit with a full width at half maximum (FWHM) of 0.56 $\pm$ 0.05 nm, and the broad PSB extending to approximately 1350 nm corresponds to a kh-oriented divacancy \cite {koehl_room_2011, falk_polytype_2013, magnusson_excitation_2018}. 

We then use the FT spectrometer to measure the spectral emission of the same PL spot (5 minutes integration time), resulting in a ZPL centered at 1075.6 $\pm$ 0.1 nm and a FWHM of 2.54 $\pm$ 0.29 nm, limited by the spectral resolution of the FT spectrometer (Fig.\ref{fig:SiC_Divacancy} (b)), which is worse than the resolution of the grating spectrometer. The center wavelength and FWHM extracted from a Lorentzian fitting are comparable to the measurements obtained with the InGaAs spectrometer. This confirms the accuracy and resolution of the FT spectrometer in identifying quantum emitters in the IR region. 
An interesting feature is that the amplitude of the phonon sideband appears to be better resolved for the FT spectrometer than for the grating spectrometer. Additional simulations presented in the Supplementary Information (Fig. S4) show that this is a consequence of the different spectral resolutions for the two instruments considered here: when increasing the spectral resolution of the FT spectrometer to the same value as the grating spectrometer, the spectra appear identical. 

We then characterize the minimum time required for the FT interferometer to resolve the ZPL of the divacancy by acquiring a series of measurements with varying integration times per step. Here, 'step' refers to the incremental adjustment of the position of the moving wedge corresponding to an incremental delay. As shown in Fig.\ref{fig:SiC_Divacancy} (e), an integration time of 10 ms per step is sufficient to resolve the ZPL and identify the kh-divacancy \cite {koehl_room_2011, falk_polytype_2013, magnusson_excitation_2018}, corresponding to a minimum measurement time of 37 seconds for 1293 motor steps. At such low integration time, the total data acquisition time is mainly limited by motor motion. 

The measurements reported in Fig.\ref{fig:SiC_Divacancy} demonstrate that the FT spectrometer serves as a valuable tool to investigate quantum emitters, particularly in the infrared region, where InGaAs spectrometers are known for their high dark noise \cite{klem_mesa-isolated_2009, fathipour_advances_2016, zhu_review_2024}. Its spectral resolution, acquisition time, and cost-effectiveness compared to InGaAs spectrometers enable the correct identification of unknown defects by effectively determining their ZPL. 

\begin{figure*}[!htbp]
\centering
\includegraphics[width=1\textwidth]{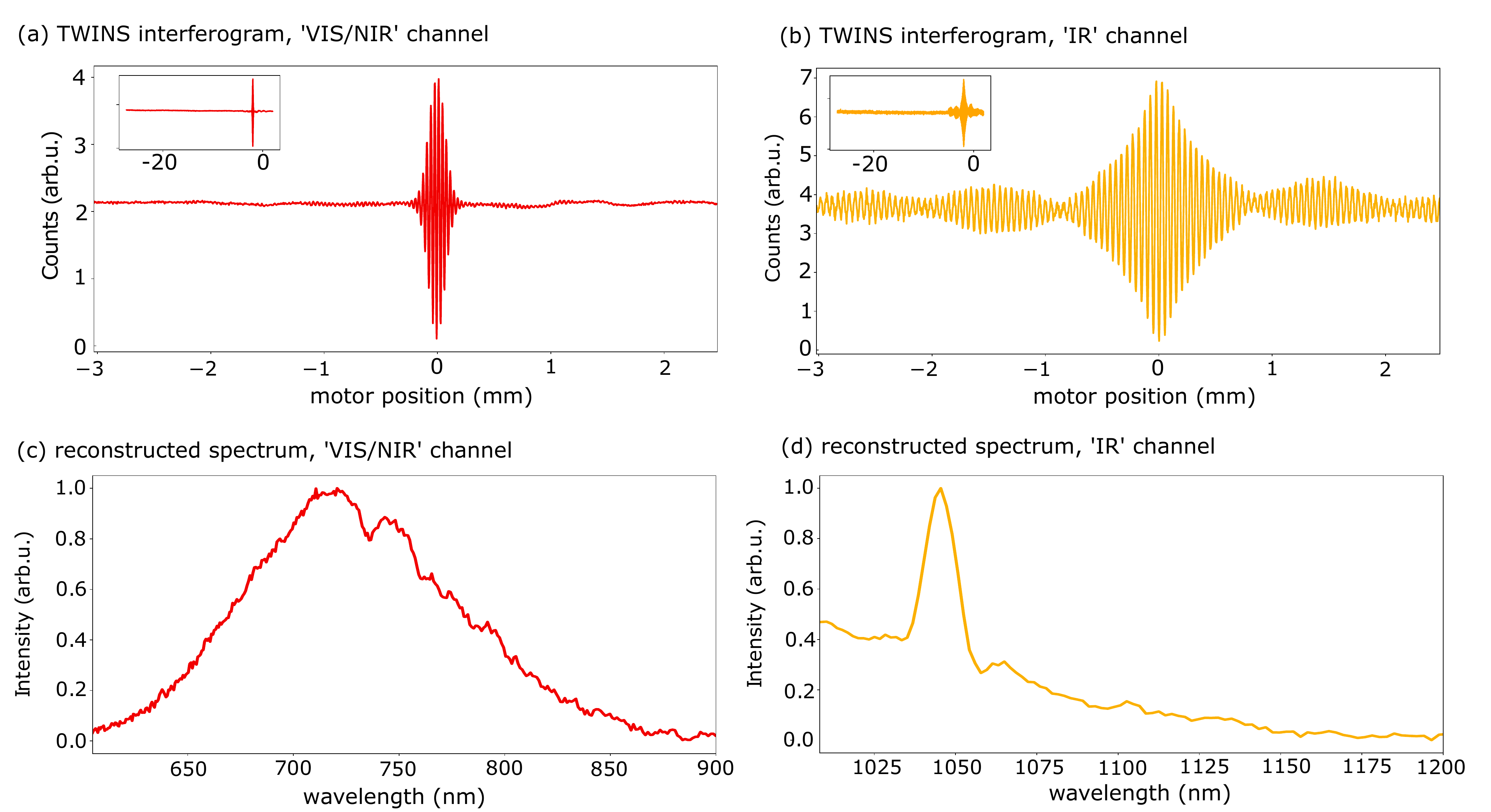}
\caption{\textbf{Demonstration of broadband PL spectroscopy of quantum emitters.} We perform FT spectroscopy of the emission from an ensemble of NV centres in diamond, detecting different wavelength ranges in parallel with two SNSPDs. \textbf{(a)} and \textbf{(b)} show respectively the interferograms on detectors optimized for the 600-900 nm range (emission from the NV triplet state) and for the 1000-1300 nm range (emission from the singlet state). The corresponding spectra are shown in \textbf{(c)} and \textbf{(d)}.}
\label{fig:NV_broadband}
\end{figure*}

\section {Broadband Fourier spectroscopy of quantum emitters}
\label {sec:broadband}

In this section, we demonstrate that the time-domain FT spectrometer can be used for broadband spectroscopy of quantum emitters when coupled in parallel to multiple single-photon detectors operating in different wavelength ranges. Throughout Sections \ref{sec:broadband} and \ref{sec:time_resolved}, we identify as ``VIS/NIR'' the spectral region between 600-1000 nm and as ``IR'' the spectral region between 1000-1200 nm. 

We utilize an ensemble of nitrogen vacancy (NV) centers in diamond.  The optical dynamics of the negatively-charged state of the NV center (NV$^{-}$) can be modeled as that of a six-level system \cite{dumeige_infrared_2019, magaletti_modelling_2024} which emits both VIS/NIR and IR photons, as shown in Fig. \ref{fig:NV_levels_setup} (a). In our model we do not distinguish between the $m_s = -1$ and $m_s = +1$ spin sub-levels in the $S=1$ triplet states: therefore the triplet ground ($^{3}$A$_2$) and excited ($^{3}$E) states only include, respectively, the levels $\lbrace \ket{g_0}, \ket{g_1} \rbrace$ and $\lbrace \ket{e_0}, \ket{e_1} \rbrace$. We also consider a singlet ground state $\ket{s_B}$ ($^{1}$E) and excited state $\ket{s_A}$ ($^{1}$A$_1$).

Green illumination with a rate  $\kappa_{exc}$ excites the system from the $^{3}$A$_2$ ground state to the $^{3}$E excited state, from which there are two possible decay pathways: First, spin-conserving transitions (rates $\kappa_{ge}^{(0)}$ and $\kappa_{ge}^{(1)}$, respectively, for spin $m_s = 0$ and $m_s = 1$) back to $^{3}$A$_2$ with emission featuring a ZPL at 637 nm and PSB extending up to 900 nm, and a lifetime of about 13 ns \cite{dumeige_infrared_2019,magaletti_modelling_2024}; Second, decay through the singlet, consisting of non-radiative decay to the singlet excited state $^{1}$A$_1$ (rates $\kappa_{es}^{(0)}$ and $\kappa_{es}^{(1)}$), followed by radiative decay to $^{1}$E with narrowband emission around 1042 nm  \cite{acosta_optical_2010, jensen_cavity-enhanced_2014, kehayias_infrared_2013, dumeige_infrared_2019} (rate $\kappa_S$), and subsequent non-radiative decay back to the triplet ground state $^{3}$A$_2$ (rates $\kappa_{sg}^{(0)}$ and $\kappa_{sg}^{(1)}$). Given that $\kappa_{es}^{(1)}$ is larger than $\kappa_{es}^{(0)}$, this second pathway is preferentially associated with $m_s = 1$.

The experimental setup is sketched in Fig. \ref{fig:NV_levels_setup} (b), and discussed in details in the SI (Section S2). For the experiments in this section, we do not use the acousto-optic modulator (AOM) path and the microwave chain depicted in Fig. \ref{fig:NV_levels_setup} (b). After the TWINS interferometer,  the PL is sent to a dichroic mirror separating the
VIS/NIR and IR components, which are detected by two separate SNSPDs optimized for the respective wavelength ranges. As the interferometer motor is stepped, one interferogram is measured on each detector, in parallel, resulting in the two spectra displayed in Fig \ref{fig:NV_broadband}.

The emission from the triplet state is expected to be about three orders of magnitude brighter than the singlet emission \cite{acosta_optical_2010}. As we are not interested here in the relative intensity of the VIS/NIR and IR emission, we simply place suitable neutral-density filters to operate both detectors at about $10^6$ cps. As expected, the detected NV emission comprises a PSB of about 200 nm in the VIS/NIR and a 1042 nm peak in the IR \cite{rogers_singlet_2015, acosta_optical_2010}. The IR spectrum also comprises a tail from the three-orders-of-magnitude stronger visible PSB. 

\begin{figure*}[!htbp]
\centering
\includegraphics[width=1\textwidth]{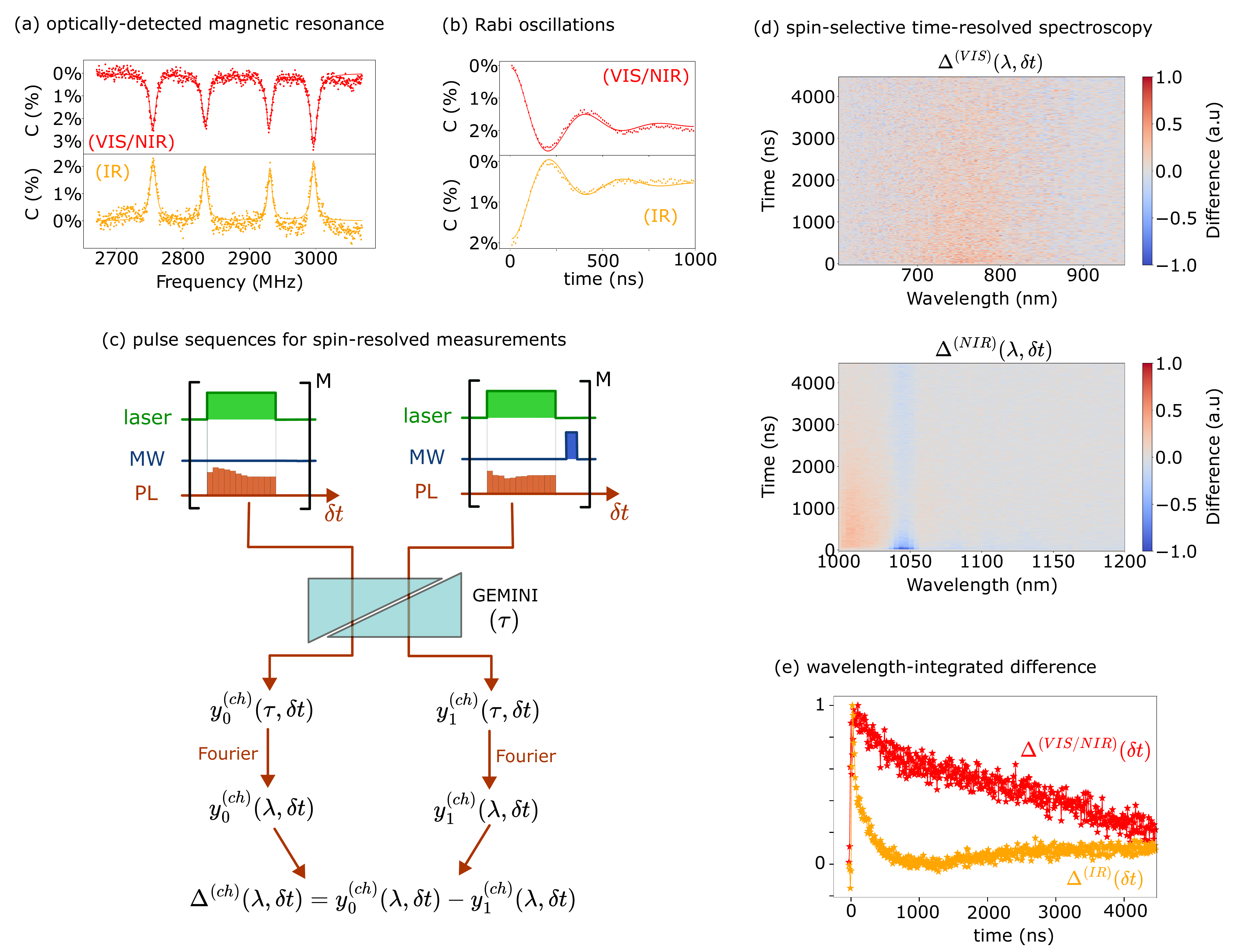}
\caption{\textbf{Spin-selective time-resolved FT spectroscopy of NV centers in diamond.} \textbf{(a)} ODMR measurement for VIS/NIR (605 nm longpass filter - 950 nm shortpass filter) and IR (1000 nm longpass filter). Optical pumping polarizes the electron spin in $m_s = 0$, which is brighter (darker) in the visible (IR). This results in PL dips (peaks) when the microwave is resonant with a spin transition.  \textbf{(b)} Rabi measurement for VIS/NIR and IR, driving the transition at 2.995 GHz with microwave pulses of different lengths. Both curves are fitted by the function $f (\tau) = A \cdot \sin(2 \tau/T_{\pi}) \cdot e^{-\frac{\tau}{T_2^*}} + B$, with the same parameters $\pi$-pulse  duration $T_{\pi}= (200.5 \pm 13.7)$ ns and $T_2^* = (245.8 ns \pm 92.5)$ ns. The relatively slow $\pi$-pulse duration is limited by the microwave power used in the experiment.
\textbf{(c)} Sketch of the pulse sequences and data analysis procedure utilized for spin-selective time-resolved measurements. We initialize the electron spin in either $m_s=0$ (left) or $m_s =1$ (right). Detecting the PL emitted during green laser excitation in both the visible (ch=VIS/NIR) and the near-infrared (ch=IR). We collect the PL photon detection times in bins $\delta t$. The PL is analyzed with the FT spectrometer, retrieving a time-resolved spectrum $y_i (\lambda, \delta t)$ for spin state $m_s = i$. To isolate the temporal dynamics from strong background from the NV ensemble, we plot the difference $\Delta (\lambda, \delta t)$.
\textbf{(d)} Experimental data for spin-selective time-resolved measurements, for the VIS/NIR (top plot, $\Delta^{(VIS/NIR)} (\lambda, \delta t)$) and the IR (bottom plot, $\Delta^{(IR)} (\lambda, \delta t)$). \textbf{(e)} Comparison between $\Delta^{(VIS/NIR)} (\delta t)$ (wavelength integrated in the range 600-900nm) and $\Delta^{(IR)} (\delta t)$ (wavelength integrated around the 1042nm peak), showing the two different timescales for the photoluminescence.}

\label{fig:Time_resolved_PL}

\end{figure*}

\section {Fourier spectroscopy of quantum emitters with nanosecond temporal resolution}
\label {sec:time_resolved}
An appealing feature of single-photon detectors is their ability to provide temporal resolution, down to few tens of picoseconds for SNSPDs. This opens the possibility of wide-band time-resolved PL spectroscopy for quantum emitters, in particular targeting different, possibly spin-dependent, emission pathways.

We demonstrate this by performing experiments on the interplay between the visible and infrared emission of an ensemble of NV centers in diamond.  As shown in Fig. ~\ref{fig:NV_levels_setup}(a) and described in Section \ref{sec:broadband}, the NV center presents a complex dynamics involving singlet and triplet excited states, with optical emission in the VIS/NIR and IR. The relevant energy levels and transitions, and the associated dynamics and rates are well known \cite{acosta_optical_2010}, enabling us to use this system to benchmark the technique.  As $\kappa_{es}^{(1)} \gg \kappa_{es}^{(0)}$, the decay through the singlet states ($\ket{s_A}$ and $\ket{s_B}$) is preferentially activated for $m_s = 1$. The longer lifetimes associated with the state $\ket{s_B}$ ($\kappa^{(0,1)}_{sg} \leq 1$ MHz) lead to a reduction in optical emission in the visible for the $m_s = 1$ spin state, compared to the $m_s = 0$ state: this spin-dependent PL intensity gives a spin readout mechanism. Furthermore, as the $\ket{s_B}$ state displays a stronger coupling to $\ket{g_0}$ than to $\ket{g_1}$ in the triplet ground state, optical excitation results in pumping into the $m_s = 0$ state, providing a spin initialization mechanism.

The optically-detected magnetic resonance (ODMR) on the NV centers is shown in Fig. ~\ref{fig:Time_resolved_PL}(a), with the red line indicating spin resonance spectrum for visible light collection, and the yellow line for infrared light collection. As the system is optically polarized in the $m_s = 0$ state, which displays higher (lower) brightness in the VIS/NIR (IR) PL, the ODMR spectrum display dips (peaks) when the microwaves are applied resonant with the transition, driving the spin into $m_s=1$.  In these measurements, a magnetic field of about 35 gauss is applied along one of the four possible NV directions in the sample by a permanent magnet. We further demonstrate spin rotations by  microwave pulses of different durations driven on the NV center spin resonance line at 2995 MHz: the resulting Rabi oscillations, in VIS/NIR (red) and IR (yellow), are reported in Fig. ~\ref{fig:Time_resolved_PL}(b). From these oscillations, we determine the optimal $\pi$-pulse duration, which is utilized in subsequent microwave pulse sequences.

To demonstrate the capability of our wide-band FT spectrometer to time-resolve spin-dependent optical dynamics with high spectral and temporal resolution, we perform pulsed experiments on the NV ensemble, starting from either $m_s = 0$ or $m_s = 1$. We expect a larger PL intensity in the VIS/NIR when the sample is polarized in $m_s =0$ compared to when it is polarized in $m_s = \pm 1$ and the opposite behavior for the IR emission \cite{dinani_bayesian_2019}, with different timescales for the dynamics resulting from the different lifetimes of the states involved.

We detect 1.6 Mcps (spectrally-integrated) for the VIS/NIR detector and 700 kcps for the IR detector, with adequate neutral density filters to avoid detector saturation. In order to compare the dynamics for the electron spin initialized in either $m_s=0$ or $m_s = 1$, we utilize two different pulse sequences. In the first one, we use a single green laser pulse of 15 $\mu$s duration, which polarizes the spin into the $m_s = 0$ state. In the second, we add a microwave $\pi$ pulse at 2995 GHz (207 ns duration, determined by the Rabi experiment) to flip the spin state from the initially-polarized $m_s=0$ state into $m_s=1$.

The dynamics is observed by detecting PL photons during the green illumination pulse, in the range 600-1200nm, as in Section \ref{sec:broadband}. For each interferometer time step $\tau_j$, we record the photon arrival times $\lbrace t_i^{(VIS/NIR)} \rbrace$ and $\lbrace t_i^{(IR)} \rbrace$ respectively, in the $VIS/NIR$ and $IR$ channels of the SNSPD system. The ultimate limit in temporal resolution is given by the detector jitter, below 50ps for our system, setting the shortest meaningful time-bin size $\delta t$. Shorter time-bins, however, reduce the number of photons in each bin, decreasing the SNR and establishing a trade-off between temporal resolution and SNR for a given integration time. For example, a brightness of 1 Mcps, corresponding to 10 cps in a 10 $\mu$s pulse, would result in $50 \times 10^{-6}$ photons in each 50ps-bin, requiring $2 \times 10^6$ pulse repetitions to achieve 100 photons per bin. 

As the ODMR dips in Fig. \ref{fig:Time_resolved_PL}(a) show a total contrast of about 2\% integrated over a large pulse duration of $>10 \mu$s, we expect very small differences in signal in each time-bin and we therefore need a sufficiently long integration time to enable such small variations to emerge above the shot noise. This leads to long integration times, so we decided to limit the temporal bin sizes to $\delta t = 10$ ns and $\delta t = 500$ps. We take 2930 time-delay ($\tau$) points for the interferogram (step-size 0.01 mm), with $1.6 \times 10^7$ repetitions for the pulse sequence for each value of $\tau$. For a pulse sequence duration of 22 $\mu s$, the whole experiment required a total integration time of 12.6 days. 

As described in Fig. \ref{fig:Time_resolved_PL} (c), the measurements output the datasets $y_0^{(ch)} (\tau, \delta t)$ and $y_1^{(ch)} (\tau, \delta t)$, describing photon counts as a function of the photon arrival time bin $\delta t$ and interferometer step $\tau$, given the spin is initialized in $m_s = 0$ and $m_s = 1$ respectively ($ch \in \lbrace VIS/NIR, IR \rbrace$). Each of the two datasets comprises data taken from the $VIS/NIR$ and $IR$ detector channels. We then Fourier-transform these with respect to the interferometer step $\tau$, retrieving the time-resolved spectra $y_0^{(ch)} (\lambda, \delta t)$ and $y_1^{(ch)} (\lambda, \delta t)$ as a function of the optical wavelength $\lambda$.

Figure ~\ref{fig:Time_resolved_PL}(d) plots the quantity $\Delta^{(ch)}(\lambda, \delta t) = \left[ y_0^{(ch)} (\lambda, \delta t) - y_1^{(ch)} (\lambda, \delta t) \right]$, i.e. the difference in counts observed between the situation with spin initially prepared in $m_s = 0$ and $m_s = 1$, respectively in VIS/NIR (top) and IR ( bottom), for a time-bin of 10 ns. We choose to plot the difference, instead of the two original datasets $y_0^{(chan)} (\lambda, \delta t)$ and $y_1^{(chan)} (\lambda, \delta t)$, as it cancels the contribution of the NVs not driven by the specific microwave tone we apply and the general background, highlighting the small difference between the two spin states.

Over the duration of the pulse, the PL reaches an identical steady-state for $m_s = 0$ and $m_s = 1$. The VIS/NIR PL in the top panel of Fig. ~\ref{fig:Time_resolved_PL} (d) shows a positive photon count difference ($y_0^{(ch)} (\lambda, \delta t) > y_1^{(ch)} (\lambda, \delta t)$), consistent with PL dips in the ODMR spectrum, with a temporal decay over a timescale of 5-7 $\mu$s. The bottom panel of Fig. ~\ref{fig:Time_resolved_PL}(d) evidences regions with positive count difference $\Delta^{(ch)}(\lambda, \delta t)$, related to the tail of the visible emission, and areas with negative $\Delta^{(ch)}(\lambda, \delta t)$, with a much shorter temporal decay, on the order of 1 $\mu$s, associated with the 1042 nm peak. The relatively slow decay time can be attributed to the low power of the green pump \cite{poulsen_optimal_2022}. This shows how spin-selective time-resolved spectroscopy enables the identification of spectral features associated with the different spin sub-levels.

\section {Discussion and conclusions}

In this work we have demonstrated the ability of a common-path birefringent interferometer to perform time-domain FT spectroscopy for the characterization of PL of quantum emitters, comparing its performance to a frequency-domain grating-based spectrometer. We have shown that the FT spectrometer can be competitive in measuring the spectrum of single quantum emitters in spectral regions where low-noise silicon cameras are not available, such as the IR (and telecom) range, with data acquisition timescales on the order of seconds to minutes. FT spectrometers are thus proposed as a lower cost alternative to grating-based spectrometers equipped with expensive liquid nitrogen cooled InGaAs cameras. SNSPDs are of course expensive as well, but any laboratory performing research on telecom-range quantum emitters will naturally have access to them. An alternative possibility could be to utilize grating-based spectrometers coupled to SNSPD arrays, which are an active area of technology development \cite{miki_64-pixel_2014, wollman_kilopixel_2019}. Such detectors are however still not widely available, and no time-resolved spectroscopy application using them has been reported to the best of our knowledge. 

The TWINS interferometer has a high throughput, due to the lack of slits, which can be further enhanced by suitable anti-reflection coatings, at the expense of broadband operation. Its spectral resolution, which depends on the maximum achievable replicas delay, can be further improved by increasing the size of the birefringent wedges or moving to materials with larger birefringence, such as $YVO_4$.

We envision several possible applications of the techniques demonstrated in this manuscript. First of all, as shown in Section \ref{sec:divacancy_experiments}, FT spectroscopy can be used to identify quantum emitters, in particular in IR (and in the telecom range) where InGaAs cameras are limited by poor performance. Examples are divacancies \cite{christle_isolated_2015, christle_isolated_2017, anderson_five-second_2022}, NV centers \cite{jiang_quantum_2023} and vanadium \cite{cilibrizzi_ultra-narrow_2023} impurities in SiC, G-centers \cite{redjem_single_2020, durand_broad_2021, prabhu_individually_2023} and T-centers \cite{inc_distributed_2024, johnston_cavity-coupled_2024} in silicon, erbium impurities \cite {yin_optical_2013, dibos_atomic_2018,  ourari_indistinguishable_2023, gritsch_optical_2024} in different host materials. It can also be used to monitor the conversion to the telecom range of the emission from systems operating in VIS/NIR, through suitable non-linear optical processes \cite{ikuta_wide-band_2011, de_greve_quantum-dot_2012, dreau_quantum_2018, arenskotter_telecom_2023}. All these examples are of crucial importance for applications requiring long-distance sharing of quantum states, such as quantum networking \cite{awschalom_quantum_2018}.

In many situations, the types of emitters which could occur in the sample are constrained by factors such as specific irradiation or annealing conditions. In such cases, where one aims, for example, only to distinguish between different types of divacancies, data acquisition times could be further decreased by compressed sensing \cite{charsley_compressive_2022}, or tailored protocols only detecting interferometer delay steps that give the most information for identifying the specific system of interest. 

As we have shown in Section \ref{sec:time_resolved}, this system can be used to investigate spin-resolved PL dynamics with high temporal resolution. While we have benchmarked this for a system, the NV center in diamond, with known transitions and dynamics, it could be helpful when characterizing different types of new quantum emitters to find the optimal ones for specific applications, and in developing optimal control pulse sequences using multi-color pulsed illumination and detection. The same concept can be applied to study the dynamics dependent on other degrees of freedom, such as the charge state for solid-state defects \cite{hopper_near-infrared-assisted_2016, wirtitsch_exploiting_2023}.

Beyond quantum technology, further applications could be found in the characterization of molecules with fast switching between different fluorescence spectra, for example photo-switchable molecules \cite{richers_coumarin-diene_2018}, solvatochromic molecules exhibiting emission wavelength shifts depending on the polarity of the solvent \cite{anandhan_solvatochromism_2019}, and spin and exciton dynamics in organic light emitting diodes \cite{scharff_complete_2021, hung_high-performance_2024}.

\section*{acknowledgments}
We express our gratitude to Sam Bayliss, Fabrizio Preda, Patrick Murton, Amit Finkler, Gianluca De Ninno and Aldona Mzyk for valuable discussions. This work is funded by the Engineering and Physical Sciences Research Council (EP/S000550/1, EP/V053779/1, EP/Z533208/1, EP/Z533191/1), the European Commission (QuanTELCO, grant agreement No 862721; QRC-4-ESP, grant agreement 101129663; QUEST, grant agreement 101156088; QuSPARC, grant agreement 101186889; QUONDENSATE, grant agreement 101130384), the Defence Science and Technology Laboratory (contract DSTL0000002448), Vinnova (project 2024-00461), and the Swedish Research Council (Grant No. VR 2020-05444).  This work is also funded by the project 23NRM04 NoQTeS, which has received funding from the European Partnership on Metrology, co-financed from the European Union’s Horizon Europe Research and Innovation Programme and by the Participating States, and by the European Union NextGenerationEU Programme with the I-PHOQS Infrastructure [IR0000016, ID D2B8D520, CUP B53C22001750006] "Integrated infrastructure initiative in Photonic and Quantum Sciences".

\clearpage

\renewcommand{\thepage}{S\arabic{page}}
\renewcommand{\thesection}{S\arabic{section}}
\renewcommand{\thetable}{S\arabic{table}}
\renewcommand{\thefigure}{S\arabic{figure}}
\renewcommand{\figurename}{Supplementary Figure}
\setcounter{page}{1}
\setcounter{figure}{0}
\setcounter{section}{0}

\onecolumngrid

\section*{Supplementary Information}

\section {Additional numerical simulations}

\begin{figure*}[!htbp]
\centering
\includegraphics[width=1\textwidth]{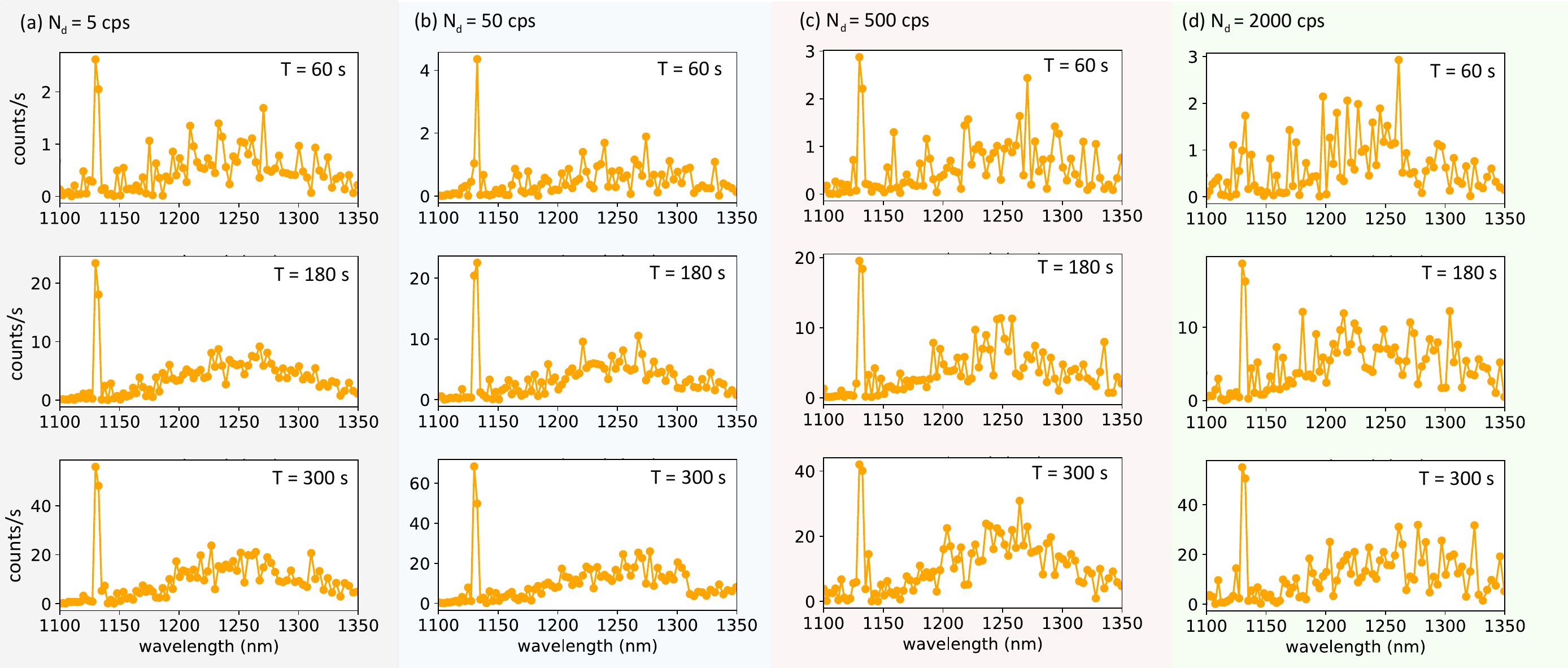}
\caption{\textbf{Numerical simulations comparing the effect of detector dark counts on the Fourier spectrometer.}  In all plots, we consider a divacancy in SiC emitting $2$k photons per second, with Debye-Waller factor $0.04$. We change the detector dark counts as \textbf{(a)} $N_d = 5$ cps, \textbf{(b)} $N_d = 50$ cps, \textbf{(c)} $N_d = 500$ cps and \textbf{(d)} $N_d = 2000$ cps. The rows correspond to different integration times (respectively $T= 60$s, $T = 180$s and $T=300$s per point). The spectrometer is quite robust against dark counts, in some conditions retrieving a meaningful spectrum even with dark counts as high as the signal counts. }
\label{fig:simulations_FT_dark_counts}
\end{figure*}

\begin{figure*}[!htbp]
\centering
\includegraphics[width=0.7\textwidth]{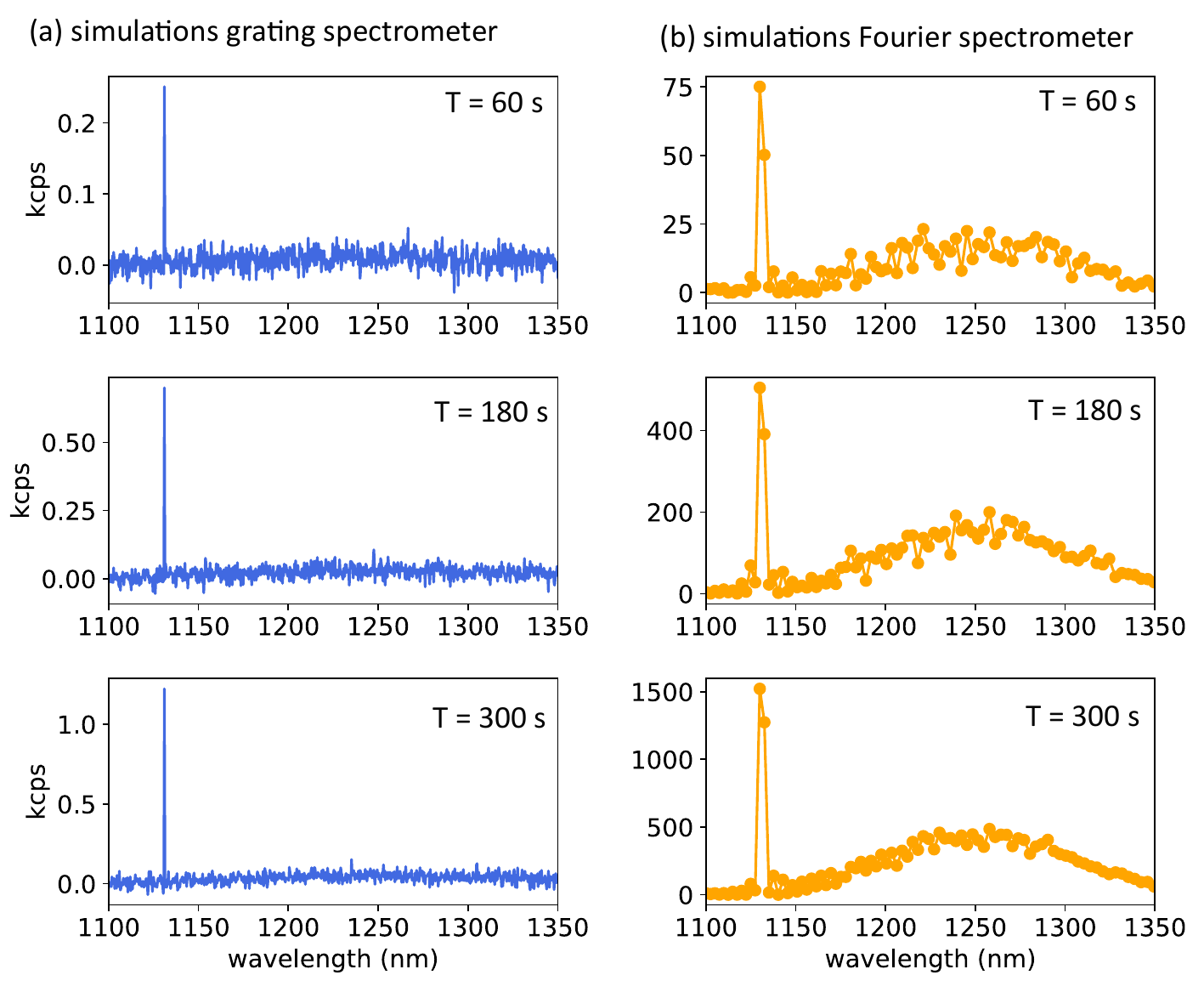}
\caption{\textbf{Numerical simulations comparing a grating and FT spectrometer.} We consider a quantum emitter with Debye-Waller factor 0.04 and brightness $10$kcps. Simulated spectra measured with the grating spectrometer coupled to a InGaAs camera (\textbf{(a)}) and the Fourier interferometer (\textbf{(b)}) described in the main text. For each instrument, we compare three total data acquisition times: $T = 60$ s, $T = 180$ s and $T = 300$ s.}
\label{fig:simulations_10kcps}
\end{figure*}

\begin{figure*}[!htbp]
\centering
\includegraphics[width=0.7\textwidth]{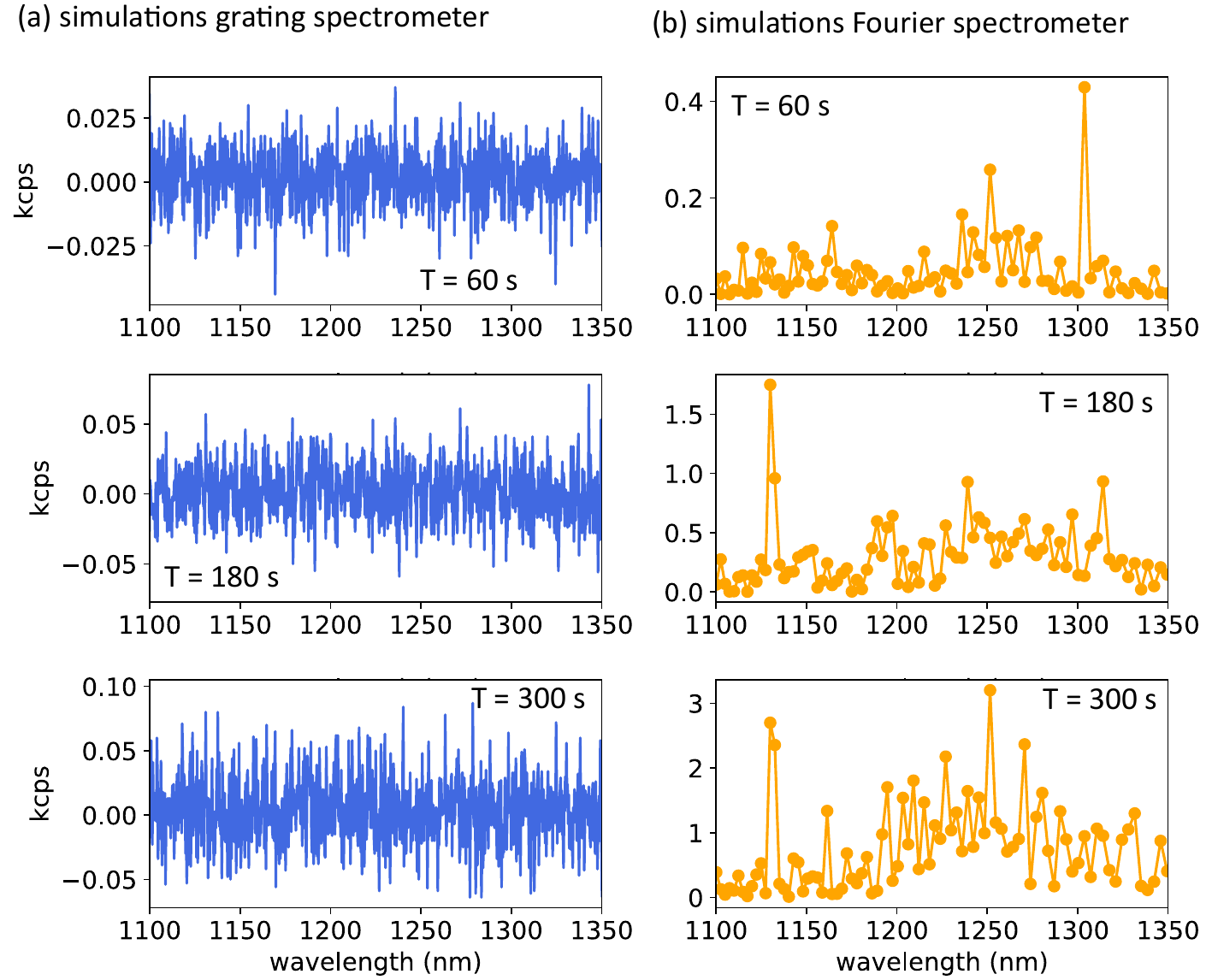}
\caption{\textbf{Numerical simulations comparing a grating and FT spectrometer.} We consider a quantum emitter with Debye-Waller factor 0.04 and brightness $500$ cps. Simulated spectra measured with the grating spectrometer coupled to a InGaAs camera (\textbf{(a)}) and the Fourier interferometer (\textbf{(b)}) described in the main text. For each instrument, we compare three total data acquisition times: $T = 60$ s, $T = 180$ s and $T = 300$ s.}
\label{fig:simulations_500cps}
\end{figure*}

\begin{figure*}[!htbp]
\centering
\includegraphics[width=1\textwidth]{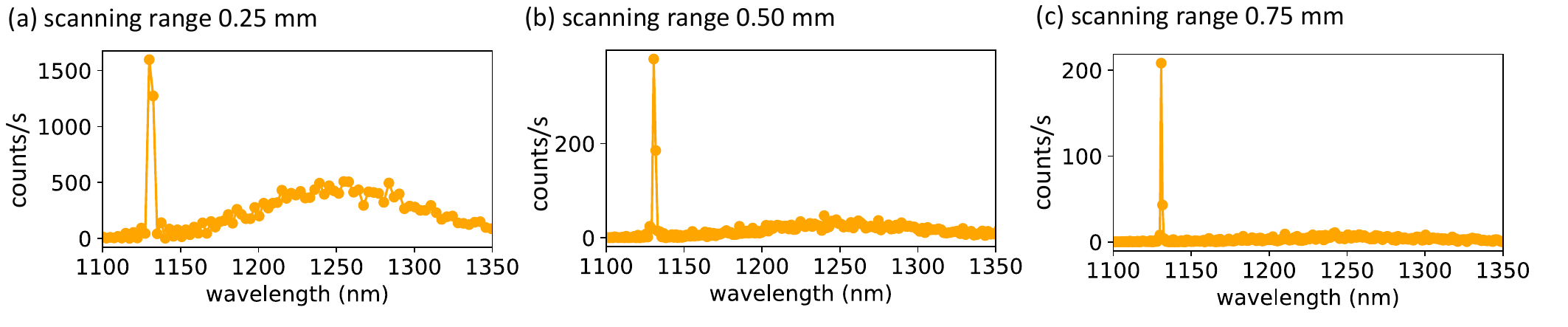}
\caption{\textbf{Numerical simulations comparing different scanning ranges for the Fourier spectrometer.} In all plots, we consider a divacancy in SiC emitting $10$k photons per second, with Debye-Waller factor $0.04$ and a detector with $50$ dark counts per second. The integration time is 0.3 seconds per point. \textbf{(a)} Simulated spectrum for a scanning range of $0.25$ mm;  \textbf{(b)} Simulated spectrum for a scanning range of $0.50$ mm; \textbf{(c)} Simulated spectrum for a scanning range of $0.75$ mm}
\label{fig:simulations_spectral_resolution}
\end{figure*}

Here we report additional numerical simulations to support the work presented in the main text. In particular:

\begin{itemize}
\item Supplementary Figure \ref{fig:simulations_FT_dark_counts} examines the role of detector dark counts in Fourier spectroscopy, showing that spectra corresponding to a single quantum emitter can be retrieved even when the noise is as intense as the signal
\item Supplementary Figure \ref{fig:simulations_10kcps} compares a grating and FT spectrometer, with the same parameters as described in the main text, when detecting a quantum emitter with brightness 10kcps.
\item Supplementary Figure \ref{fig:simulations_500cps} compares a grating and FT spectrometer, with the same parameters as described in the main text, when detecting a quantum emitter with brightness 500 cps. In this case, the Fourier spectrometer can retrieve a spectrum, while the grating-based spectrometer cannot.

\item Supplementary Figure \ref{fig:simulations_spectral_resolution} compares spectra for a single quantum emitter taken on the Fourier spectrometer with different spectral resolutions (larger delay ranges between the waveform replicas result in smaller spectral resolution).
\end{itemize}

\section {Experimental setup and samples for the experiments on NV centres in diamond}

\textbf{Sample.} We utilize an ensemble of NV centers in diamond, created by electron irradiation and annealing of a high-pressure high-temperature diamond sample purchased from Element Six Ltd. After electron irradiation by Synergy Health plc, the diamond was annealed while buried under diamond grit in a tube furnace with dry nitrogen flowing over it, following the recipe of Chu et al \cite{chu_coherent_2014}: 4 hours at 400 $^{\circ}$C, and then 2 hours at 800 $^{\circ}$C, and then 2 hours at 1200 $^{\circ}$C. After annealing, the sample was cleaned in acid. Continuous-wave electron paramagnetic resonance (EPR) was used to quantify the defect concentrations, measuring nitrogen concentration as 127 ppm and the NV$^{-}$ concentration as 5.6 ppm.

\textbf{Experimental setup. } The experimental setup is sketched in Fig. 3 in the main text. The sample is kept at room temperature, and investigated with a home-built confocal microscope featuring a high numerical aperture objective (Olympus MPlanFL N 100x, NA=0.9). With an NV concentration of 5.6ppm, corresponding to $9.9 \times 10^{-6}$ per $\mu$m$^3$, we expect about $10^{5} $ NV centers in the confocal spot. The sample is placed on motorized stages (Thorlabs Nanomax300) to enable access to different areas.

For the experiments in this section, the sample is excited by a 532 nm continuous-wave (CW) laser (CNI MGL-III-532-300mW), and the PL is selected with a dichroic mirror (`DC-1' in the figure, Semrock DiO2-R561). The PL goes through the TWINS interferometer and the output is then separated in two different wavelength ranges by a second dichroic (`DC-2', Thorlabs DMSP950R ), each directed to separate SNSPD channels. The wavelength range (605 nm - 950 nm) is sent to a channel optimised for 780 nm, while the NIR range (1000 nm - 1200 nm) is directed to a second channel optimised for the 900-1300 nm range (through two long-pass filters with cut-off at 950 nm and one long-pass filter with cut-off at 1000 nm). 

Spin initialisation, control and readout is performed by a combination of optical and microwave pulses. Microwave pulses are created by a radio-frequency switch (Minicircuits ZASW-2-50DRA+, 20 ns switching speed) acting on the continuous-wave tone from a local oscillator (R\&S SMBV100A, power 25 dBm). The pulses are amplified (Minicircuits ZHL-16W-43-S+) and delivered to the sample through a thin copper wire, creating a microwave-frequency magnetic field. A continuous-wave green laser beam (20 mW on the sample) is pulsed by an acousto-optic modulator (Isomet 556F-3), in double-pass configuration to achieve an extinction ratio of $10^5:1$.

\end{document}